\documentclass[aip,reprint]{revtex4-1}
\draft 
\usepackage{graphicx}
\usepackage{dcolumn}
\usepackage{subfigure}
\hfuzz=\maxdimen
\begin{document}

\title{Spatial dynamics analysis of polarized atom vapor} 


\author{X.Y.Hu}
\affiliation{School of Instrumentation Science and Opto-Electronics Engineering, Beihang University, Beijing 100191,
	China.}
\author{H.F.Dong}
\affiliation{School of Instrumentation Science and Opto-Electronics Engineering, Beihang University, Beijing 100191,
	China.}
\author{H.C.Huang}
\affiliation{School of Instrumentation Science and Opto-Electronics Engineering, Beihang University, Beijing 100191,
	China.}
\author{L.Chen}
\affiliation{School of Instrumentation Science and Opto-Electronics Engineering, Beihang University, Beijing 100191,
	China.}
\author{Y.Gao}
\affiliation{School of Instrumentation Science and Opto-Electronics Engineering, Beihang University, Beijing 100191,
	China.}


\date{\today}

\begin{abstract}
We analyze the spatial dynamics of polarized atom vapor and present a mathematical method	to eliminate the diffusion effect partially. It is found that the diffusion effect of polarized	atoms can be regarded as a low pass filter in spatial frequency domain and fits well with a	Butterworth filter. The fitted spatial filter can be used to restore the original magnetic image	before being blurred by the diffusion, thus improving the magnetic spatial resolution. The	results of spatial dynamics simulation and magnetic image restoration show the potential usage of this method in magnetic gradiometer and atomic magnetic microscopy.
\end{abstract}

\maketitle 

\section{\label{sec:level1}Introduction}
The temporal and spatial dynamics of polarized atom vapor is described by Bloch equation\cite{article}, which is the basic model in fields such as nuclear magnetic resonance(NMR)\cite{Hinshaw1983An,Lehmberg1972Modification}, atomic magnetometer\cite{Savukov2015Gradient,Allred2002High,doi:10.1063/1.4902567} and atomic
gyroscopes\cite{Dong2011Analysis,Kornack2005Nuclear}. In these applications the temporal dynamics of polarized atom vapor is analyzed theoretically and	verified experimentally. As most of the applications measure the average polarization of atoms in the vapor cell,
the spatial dynamics is usually ignored in the modeling	and analysis. In the case of gradiometer or array measurements, the diffusion is thought as the essential limit	for the spatial resolution\cite{Kominis2003A,Giel2000Diffusion,Kim2014Multi}.

Inspired by the work of D. Giel et al, who pointed out
that the space-time evolution of polarization can be expanded
in terms of spatial periodic functions\cite{Giel2000Diffusion}, we introduce
the spatial frequency response of the input magnetic
field, which can describe the spatial dynamics of polarized
atom vapor in nonuniform magnetic field. MATLAB
Simulink is used to simulate the polarized atom vapor
system. By setting the magnetic field distribution as one
dimensional (1D) sinusoidal waves with different spatial
frequency, we get the evolution of the atom polarization
in both time and space domain. The result shows that
the response decreases when spatial frequency of magnetic
field increases, just like a low pass spatial filter,
which can be fitted well to a Butterworth filter. Assuming
the diffusion effect is isotropy, it can be expanded
directly to a two dimensional (2D) spatial filter. By reversing
the 2D filter, we obtain a 2D high pass filter that
can be used to eliminate diffusion effect and restore the
original image partially.

The paper is organized as follows: After a brief introduction
of background and motivation in section I,
Section II describes the model and parameters calculation
used in the simulation. Section III illustrates and
discusses the results of spatial dynamics simulation. Finally,
conclusions are summarized in section IV.
\section{\label{sec:level2}MODELING AND PARAMETERS}
The Bloch equation can be written with diffusion term
as following\cite{PhysRevA.77.033408,PhysRev.104.563}:
\begin{equation}
\frac{\partial}{\partial t}\overrightarrow{P}=D\nabla^{2}\overrightarrow{P}+\gamma\overrightarrow{B}\times\overrightarrow{P}+R_{p}(s-\overrightarrow{P})-\frac{\overrightarrow{P}}{T_1,T_2}\label{eq:eq1}
\end{equation}
where \overrightarrow{P} is the polarization of alkali atoms, D is the
diffusion coefficient, $\gamma$
 is the gyromagnetic ratio, \overrightarrow{B} is the
magnetic field, $R_p$ is the pumping rate and $T_1$ and $T_2$ are
the relaxation times for polarization components parallel
and transverse to \overrightarrow{B}, respectively. The four terms on the
right-hand side describe diffusion, precession, pumping
and relaxation, respectively.

To simplify Eq\ref{eq:eq1} and obtain the numerical result of
spatial dynamics, we assume the pumping beam and the
probing beam are along z axis and x axis, respectively,
and the direction of the magnetic field is along y axis.
Besides, we also suppose that atoms are fully polarized
using a high power short pulse beam. The polarization
vector precesses freely in x$-$z plane after the pumping
pulse. In this condition, $P_y$, $T_1$ and $R_p$ can be ignored
and Eq.(\ref{eq:eq1}) can be simplified as below:
\begin{equation}
\frac{\partial}{\partial t}
\left[
\begin{array}{c}
P_x\\
P_z\\
\end{array}
\right]=D\nabla^{2}\left[\begin{array}{c}
P_x\\
P_z\\
\end{array}\right]+\gamma{B_y}\left[\begin{array}{c}
-P_z\\
P_x\\
\end{array}\right]-R_r\left[\begin{array}{c}
P_x\\
P_z\\
\end{array}\right]
\label{eq:eq2}
\end{equation}
where relaxation rate $R_r = 1/T_2$ and diffusion coefficient
D can be calculated according to the experimental
setup.

Considering that the atom vapor cell can be antirelaxation
coated, or buffered with high pressure gas and
can work under spin-exchange relaxation free (SERF)
regime, wall collision relaxation and spin-exchange relaxation
can be neglected. Moreover, as we measure the
local field instead of the average field in the vapor cell,
gradient broadening can also be ignored. So the relaxation
rate $R_r$ is mainly decided by the spin destruction\cite{Kornack2006A},
\begin{equation}
R_r \approx R_sd =\bar{\upsilon}_\alpha\sigma^{sd}_\alpha n_\alpha+\bar{\upsilon}_q\sigma^{sd}_q n_q+\bar{\upsilon}_b\sigma^{sd}_b n_b\label{eq:eq3}
\end{equation}
where the first term denotes the collision between alkali
atoms themselves, the second term denotes the quench collision and the third term denotes the collision between alkali atom and buffer gas atom. $\bar{\upsilon},\sigma^{sd}$ and n are the relative velocity, collision cross-section of collision pair
and density of atoms, respectively. The subscripts $\alpha$, \emph{q}
and b are for alkali atoms, quenching gas atoms and buffer
gas atoms, respectively.
The diffusion coefficient depends on the temperature
and the pressure of the gas\cite{Giel2000Diffusion},
\begin{equation}
D=D_0\left(\frac{P_0}{P}\right)\left(\frac{T}{T_0}\right)^{3/2}\label{eq:eq4}
\end{equation}

where $T_0 = 273.15K$ is the standard temperature,
$P_0 = 760Torr$ is the standard pressure and $D_0$ is the
standard diffusion coefficient at $T_0$ and $P_0$. Standard
diffusion coefficients of Cesium atom in He and $N_2$ are
$0.39cm^2/s$\cite{Ishikawa:99} and $0.087(15)cm^2/s$\cite{PhysRevA.5.993.3}, respectively, which
are used in our choice of D in section III.
\section{\label{sec:level3} NUMERICAL SIMULATION AND RESULTS ANALYSIS}
\subsection{\label{sec:level4}Spatial frequency response simulation}
In the simulation, $R_r = 300s^{-1}$ and D = 0$\sim$1$cm^2/s$
are choosed according to section II with typical parameters
in the atom vapor polarization experiments. And
to obtain the spatial frequency response we set $B_y$ as a
magnetic field with spatial sinusoidal distribution on z
axis and $B_x = B_z = 0$. Thus the input magnetic field
can be written as
\begin{equation}
B_y(z)=B_0\sin(\omega z)\label{eq:eq5}
\end{equation}

where $B_0$ is the magnetic amplitude, $\omega$ is the spatial
angular frequency and z is the spatial position along z
axis.

By simulating the model in Eq.(\ref{eq:eq2}) with diffusion
term, we can get the temporal and spatial polarization
$\hat P_x(z,t_d)$ . Then the output magnetic field can be calculated
by,

\begin{equation}
\hat B_y(z)=\frac{\arcsin(-\hat{P}_x(z,t_d)\emph{e}^{R_rt_d})}{2\pi\gamma t_d}\label{eq:eq6}
\end{equation}
pulse pumping magnetometer for different D and $\omega$.
Time delay $t_d = 10\mu s$ and relaxation rate $R_r = 300s^{-1}$.

\begin{figure}
	\includegraphics[width=3.2in]{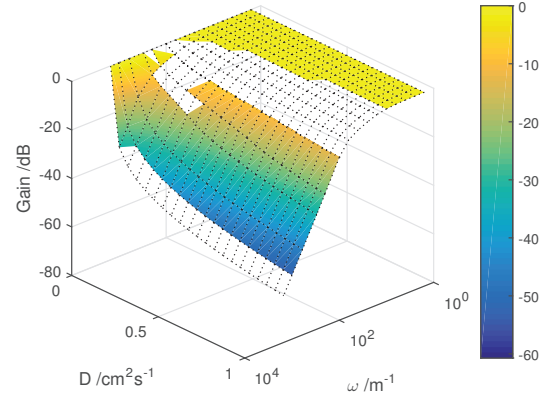}
	\caption{\label{fig:1} (Color Online) Amplitude response of the short
		pulse pumping magnetometer for different D and $\omega$.
		Time delay $t_d$=10s and relaxation rate $R_r=300s^{-1}$.}
\end{figure}

According to our simulation, $\hat{B}_y(z,t_d)$ also follows the
sinusoidal distribution with the same spatial frequency,
i.e., $\hat{B}_y(z)\approx\hat{B}_0\sin(\omega z)$.We get the corresponding magnetic
amplitude $\hat{B}_0$ of different $\omega$, $t_d$ and D. With a
certain time delay $t_d$ the amplitude magnification ${\hat{B}_0}/{B_0}$
decreases when spatial angular frequency $\omega$ increases, as
shown in Fig.\ref{fig:1}.
The figure also shows that the cutoff spatial
frequency of the system decreases with the increase
of diffusion coefficient.

\subsection{Reverse approximation of diffusion process}
\begin{figure}[ht]
	\centering
	\subfigure[]{\includegraphics[width=1.6in]{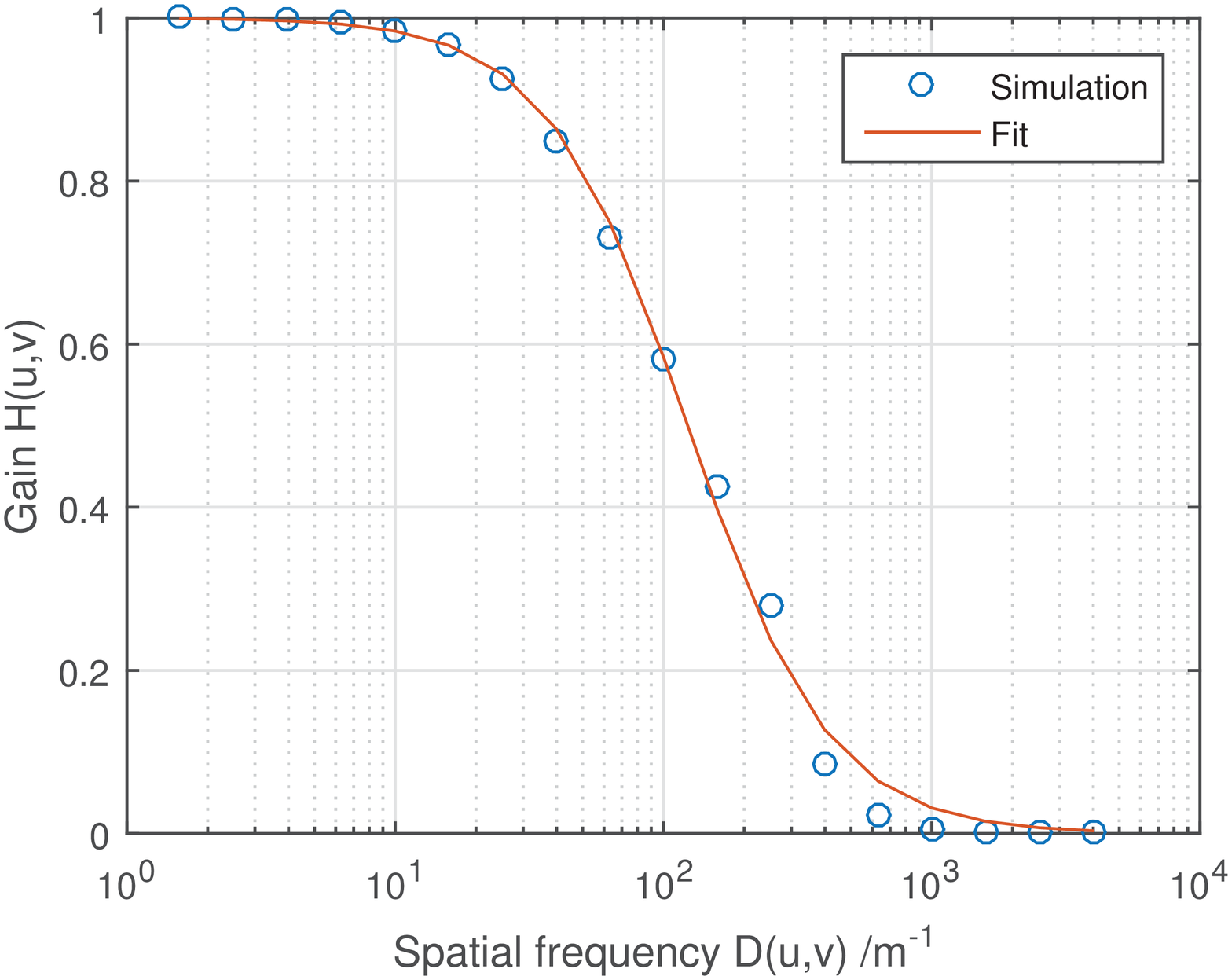}
		\label{fig:subfigure1}}	
	\subfigure[]{\includegraphics[width=1.6in]{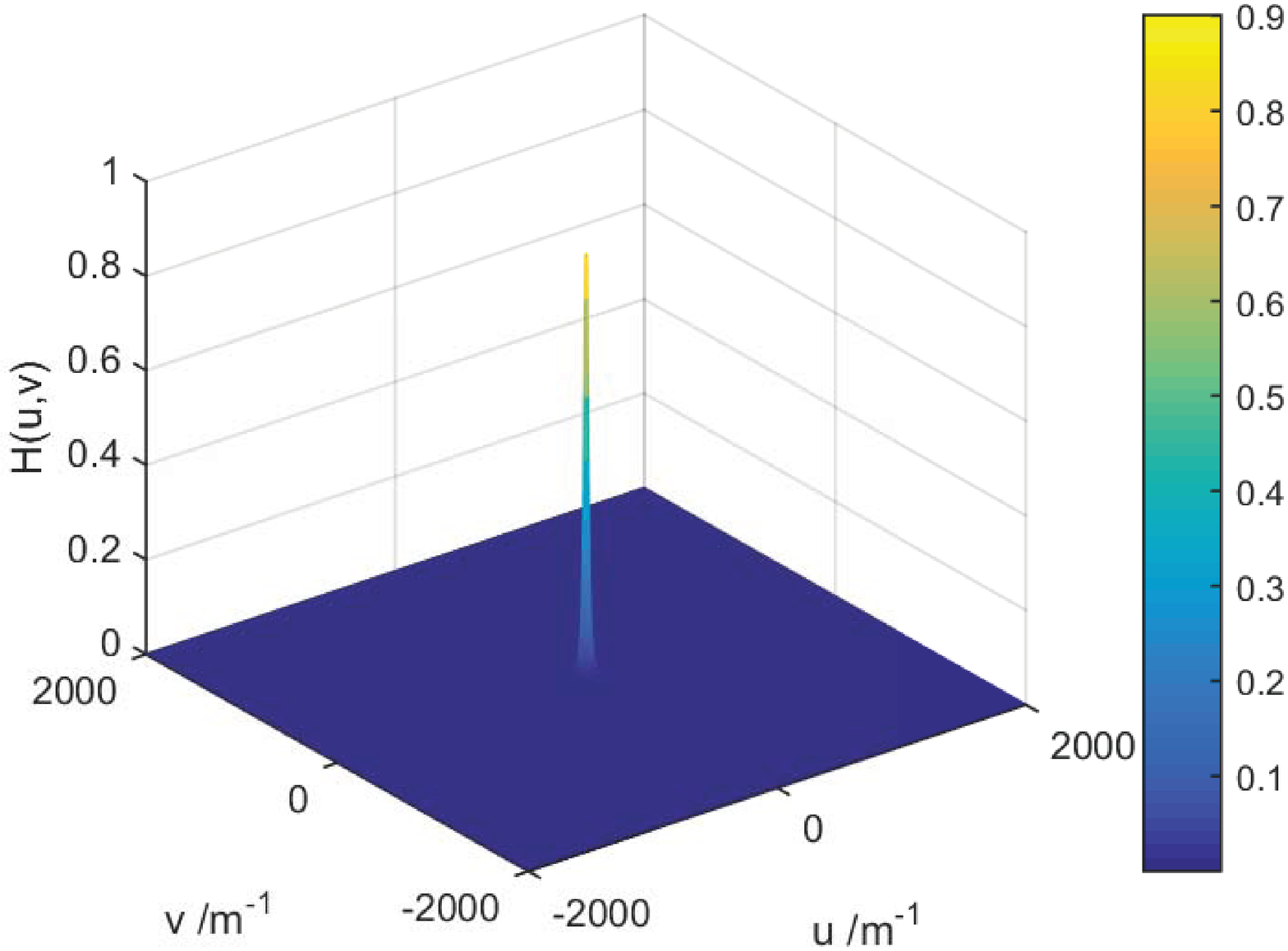}
		\label{fig:subfigure2}}
	\label{fig:2}
	\caption{(Color Online) (a) The spatial frequency response when $t_d$=10s and D = $0.6cm^2/s$. (b) The plot of the 2D filter generated by the fit curve in (a).}
\end{figure}

Fig.\ref{fig:subfigure1} 
is the spatial response when $t_d=\rm10\mu s$ and $D=\rm0.6cm^2/s$, which can be fitted well with a Butterworth filter $1/\sqrt{1+(\omega/\omega_0)^{2n}}$. The phase response of Butterworth filter is ignored due to the isotropy of diffusion. The corresponding 2D Butterworth filter can be expressed as
\begin{equation}
H(u,v) = \frac{1}{1+(D(u,v)/D_0)^{2n}}
\label{eq:eq7}  
\end{equation}

where $D(u,v)$ is the distance between a point $(u,v)$ in the frequency domain and the center of the frequency rectangle, and $D_0$ is the cutoff frequency\cite{gonzalez2009digital}. Fig.~\ref{fig:subfigure2} illustrates the corresponding 2D spatial filter, which is expanded from the fitted curve of Fig.~\ref{fig:subfigure1}.

As the 2D filter in Fig.~\ref{fig:subfigure2} is generated by the spatial frequency response of Bloch equation simulation, it approximately represents the diffusion effect. Furthermore, we can restore the original magnetic image with a reversed 2D filter $1-H(u,v)$ in Fig.~\ref{fig:subfigure2}.

\begin{figure}[ht]
	\centering
	\subfigure[]{\includegraphics[width=1.6in]{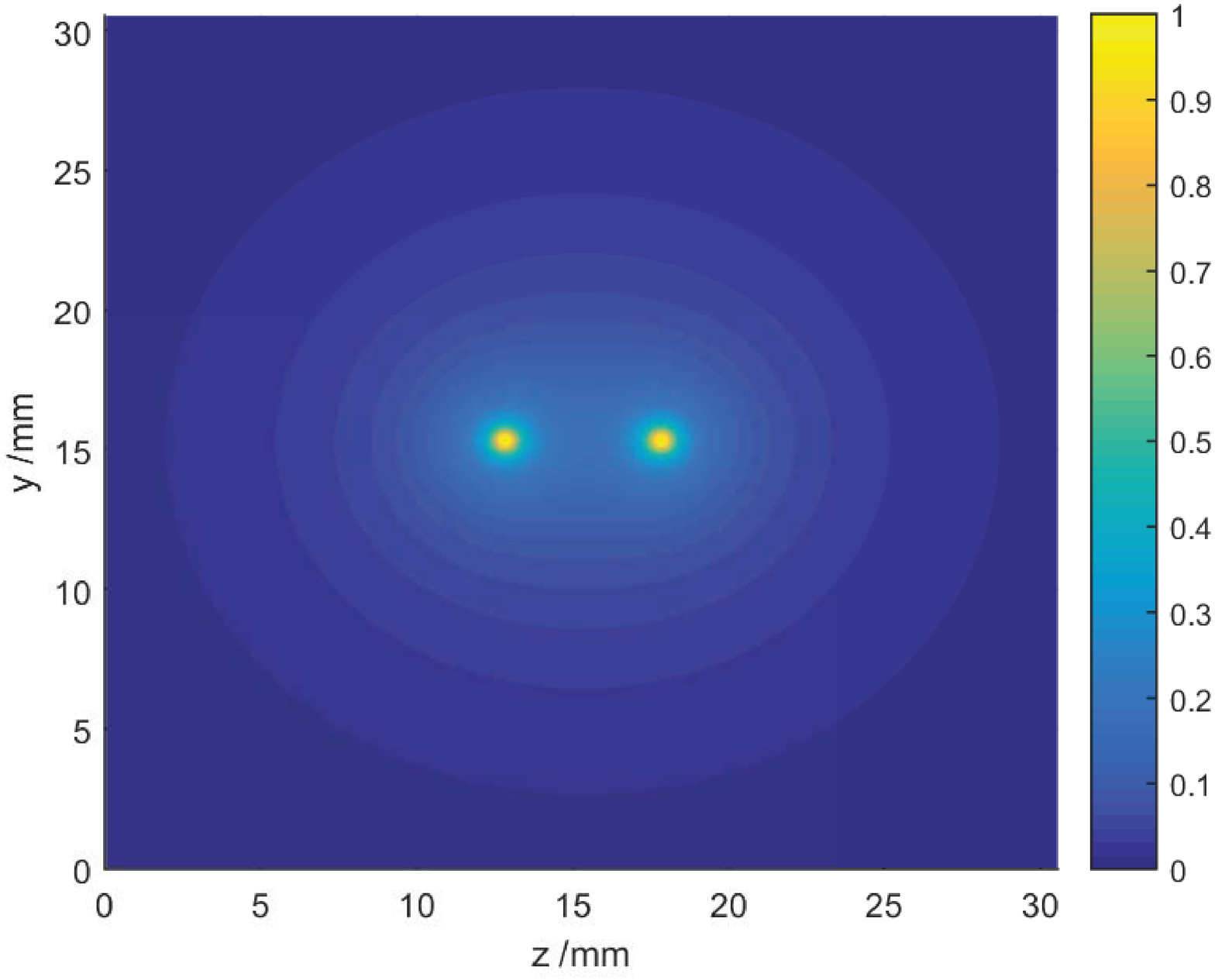}
		\label{fig:subfigure3}}	
	\subfigure[]{\includegraphics[width=1.6in]{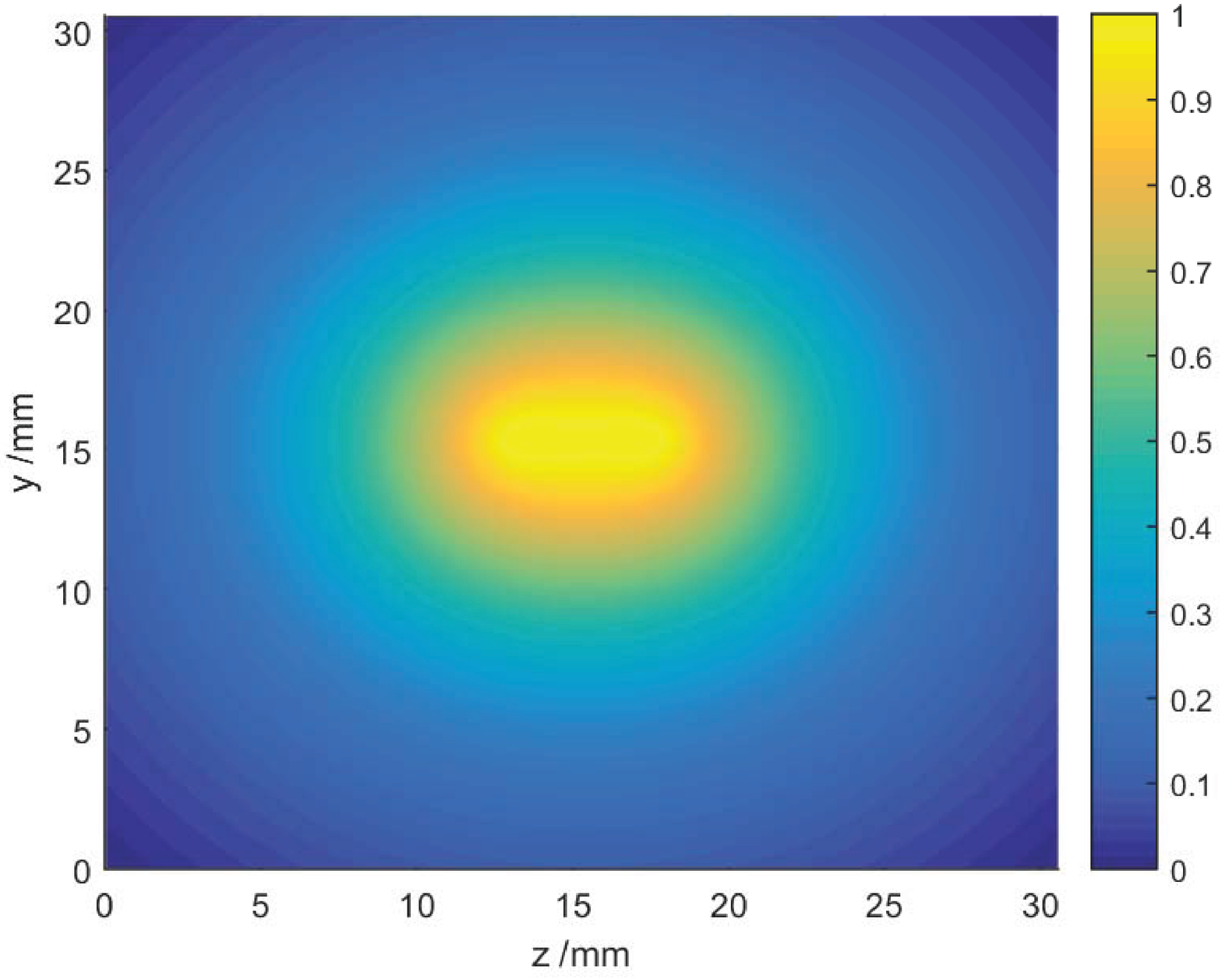}
		\label{fig:subfigure4}}
	\subfigure[]{\includegraphics[width=1.6in]{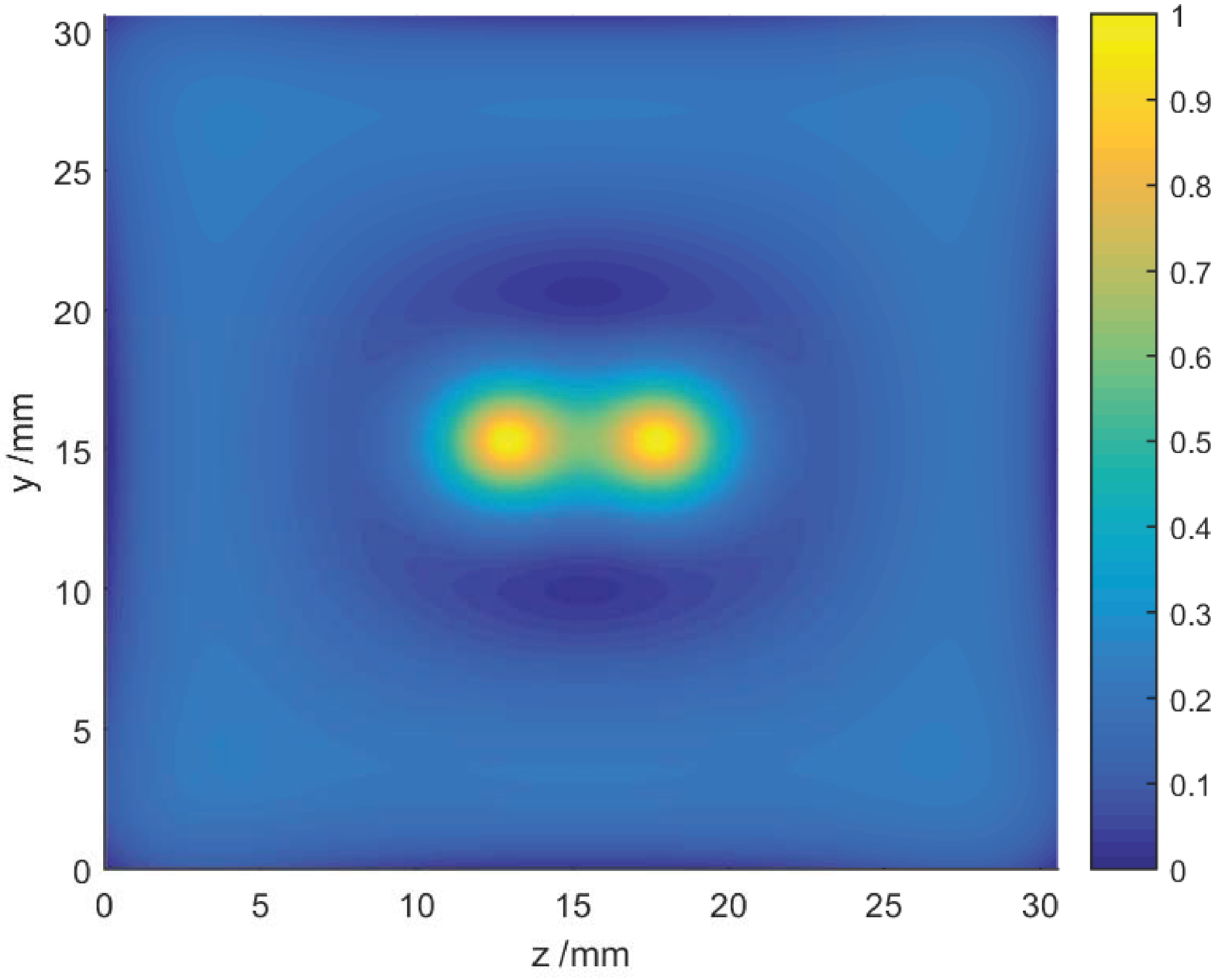}
		\label{fig:subfigure5}}
	\subfigure[]{\includegraphics[width=1.6in]{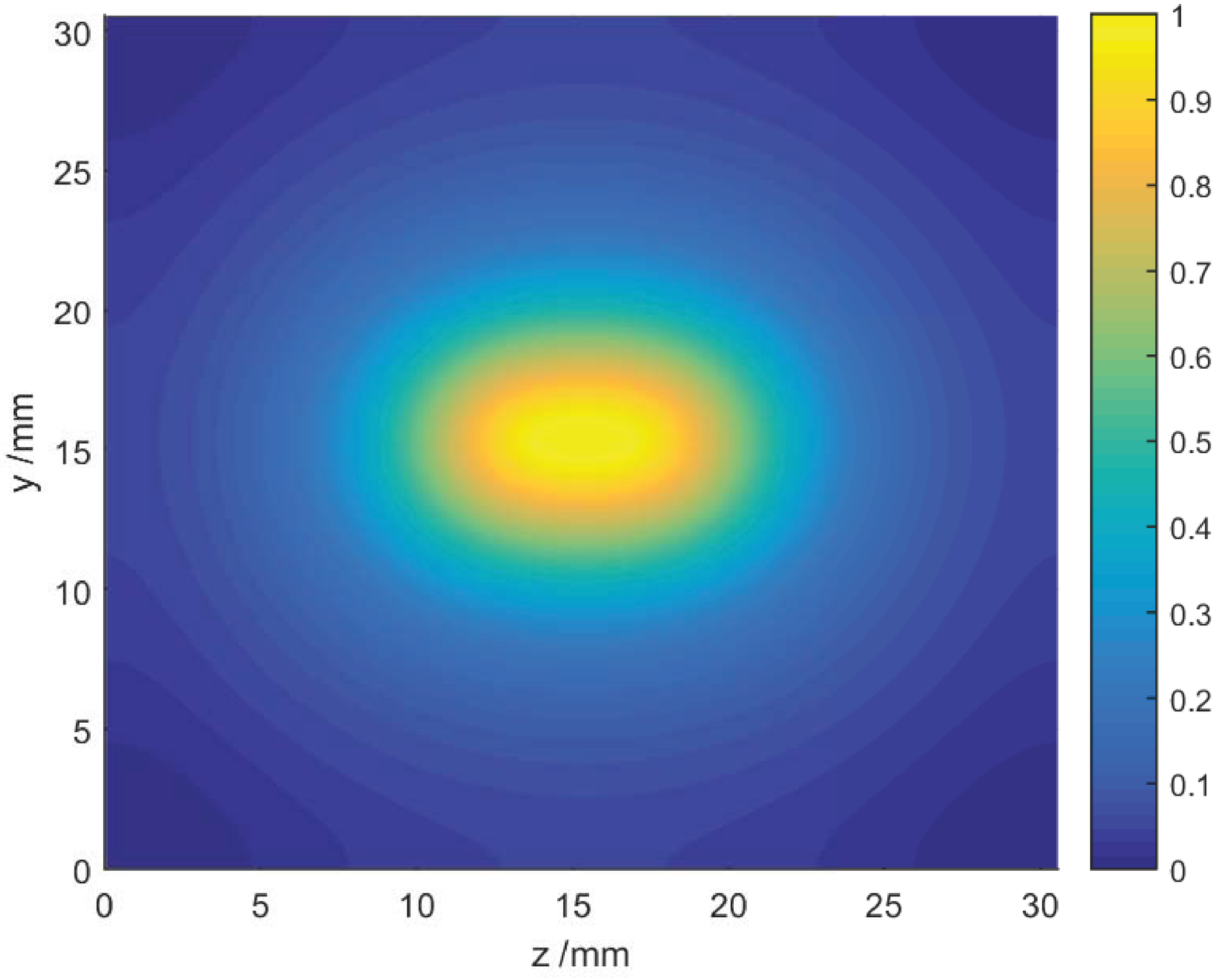}
		\label{fig:subfigure6}}
	\label{fig:3}
	\caption{(Color Online)
		(a) The original magnetic field of two dipoles in $y-z$ plane. The distance between the two dipoles is $\rm5mm$.
		(b) The measured magnetic field, where the dipoles can hardly be distinguished due to diffusion. This image is calculated using the simulation result of $P_x$ and Eq.(\ref{eq:eq6}).
		(c) The restored magnetic field using reversed 2D filter $1-H(u,v)$.
		(d) Filtered image from (a), with filter $H(u,v)$ applied. 
		To provide a better view, all of these images are normalized.}
\end{figure}

To provide a better view of the restoration effect, we simulate the atom vapor system in magnetic field of two close magnetic dipoles, and assume the dipoles is small enough to ignore the magnetic field inside the dipoles. The magnetic field $B_y$ vaires inversely with the third power of distance in $y-z$ plane, as show in  Fig.\ref{fig:subfigure3} . 
The measured magnetic image simulated using Bloch model, i.e. Eq.(\ref{eq:eq2}), and 2D filter, i.e. Eq.(\ref{eq:eq7}), are displayed in Fig.~\ref{fig:subfigure4} and Fig.~\ref{fig:subfigure6}, respectively, where the two dipoles can hardly be distinguished. With a reversed 2D filter $1-H(u,v)$, we get the restored image from Fig.~\ref{eq:eq4} and the result is shown in Fig.~\ref{fig:subfigure5}. In the restored image, the two dipoles can be distinguished again and thus the spatial resolution is improved. The slightly diffrence between Fig.~\ref{fig:subfigure3} and Fig.~\ref{fig:subfigure5} may be due to the fit error in the high frequency part. 

\section{Conclusion} 
\label{sec:conclusion}
In summary, the spatial dynamics of polarized atom vapor is analyzed based on the Bloch equation and short pulse pumping and probe scheme. The simulated spatial frequency response fits well with a low pass Butterworth filter. By passing the magnetic image through the reversed spatial fitter, we eliminate partially the diffusion effect and increase the spatial resolution of the image. This analysis and restoration method can be used in spatial magnetometry, such as magnetic gradiometer and atomic magnetic microscopy.

\section*{Acknowledgement}
The authors thank the support by National Natural Science Foundation of China under Grant No. 61074171 and 61273067 and National Program on Key Basic Research Project of China (2012CB934104). The authors would like to thank Dr. Iannis K. Kominis for the beneficial discussion on spatial resolution, which is another great inspiration to the idea of this paper.
\section*{References}
\bibliography{xuyang}

\end{document}